\documentclass[11pt]{article}
\usepackage{moriond,epsfig}

\def\be{\begin{equation}}
\def\ee{\end{equation}}
\def\bea{\begin{eqnarray}}
\def\eea{\end{eqnarray}}

\begin{document}
\vspace*{4cm}
\title{TOP MASS EXTRACTION FROM CROSS SECTION MEASUREMENTS}

\author{ S. CHEVALIER-THERY }

\address{on behalf of the D0 collaboration\\
LPTHE, 4 place Jussieu, \\
75252 Paris Cedex 05, France\\
IRFU/SPP, B\^{a}timent 141, CEA SACLAY \\
91191 Gif-Sur-Yvette Cedex, France}

\maketitle\abstracts{
An extraction of the top mass from cross section measurements in the dilepton and lepton+jets channels with an integrated luminosity of about $1\ fb^{-1}$ is given. Two ways are presented : one based on the simple lecture of the graph of the cross section as a function of the top mass and the other based on the use of probability density functions. The final result is consistent with direct measurements but with a greater uncertainty.
}
\section{Introduction}

At the Tevatron, the top quark mass is measured directly in three decay channels~\footnote{In the Standard Model, the branching ratio $ t \rightarrow Wb$ is close to $100\%$ so the channels are classified according to the decay of the W.} : the dilepton channel, the lepton+jets channel and the all-jets channel. The world average based on the combination of the measurements from the D0 and CDF collaboration~\cite{un} is $M_{t}=172.6\pm 1.4~\rm{GeV}$. Although the precision of this measurement is high, the mass extracted depends heavily on Monte Carlo simulation, in particular Pythia, and is determined in a not well-defined scheme of renormalization, nor at a well-precised order of perturbation. So to determine the top mass, in a well-defined theoretical framework, an alternative way is to extract it from cross section
measurements and theoretical predictions in NLO QCD, including 
resummations in higher orders QCD.

\section{Extraction of the top mass and combination of the uncertainties}

\subsection{First approach}

  Information about the top mass can be gained from
cross section measurements. These analyses have the advantage of being
simple and not relying on the simulation of the signal except for the
determination of signal detection efficiencies.

  From the intersection of the measured curves with the theoretical
predictions the measured top quark mass can be extracted (see figure 1).
The uncertainty on the central value is then simply given by the
intersection of the two uncertainty bands, assuming that experimental
uncertainties are not correlated with the theoretical
errors on the signal.

\begin{figure}[h!]
\centering
%\rule{5cm}{0.2mm}\hfill\rule{5cm}{0.2mm}
%\vskip 2.5cm
%\rule{5cm}{0.2mm}\hfill\rule{5cm}{0.2mm}

 \includegraphics[height=2in]{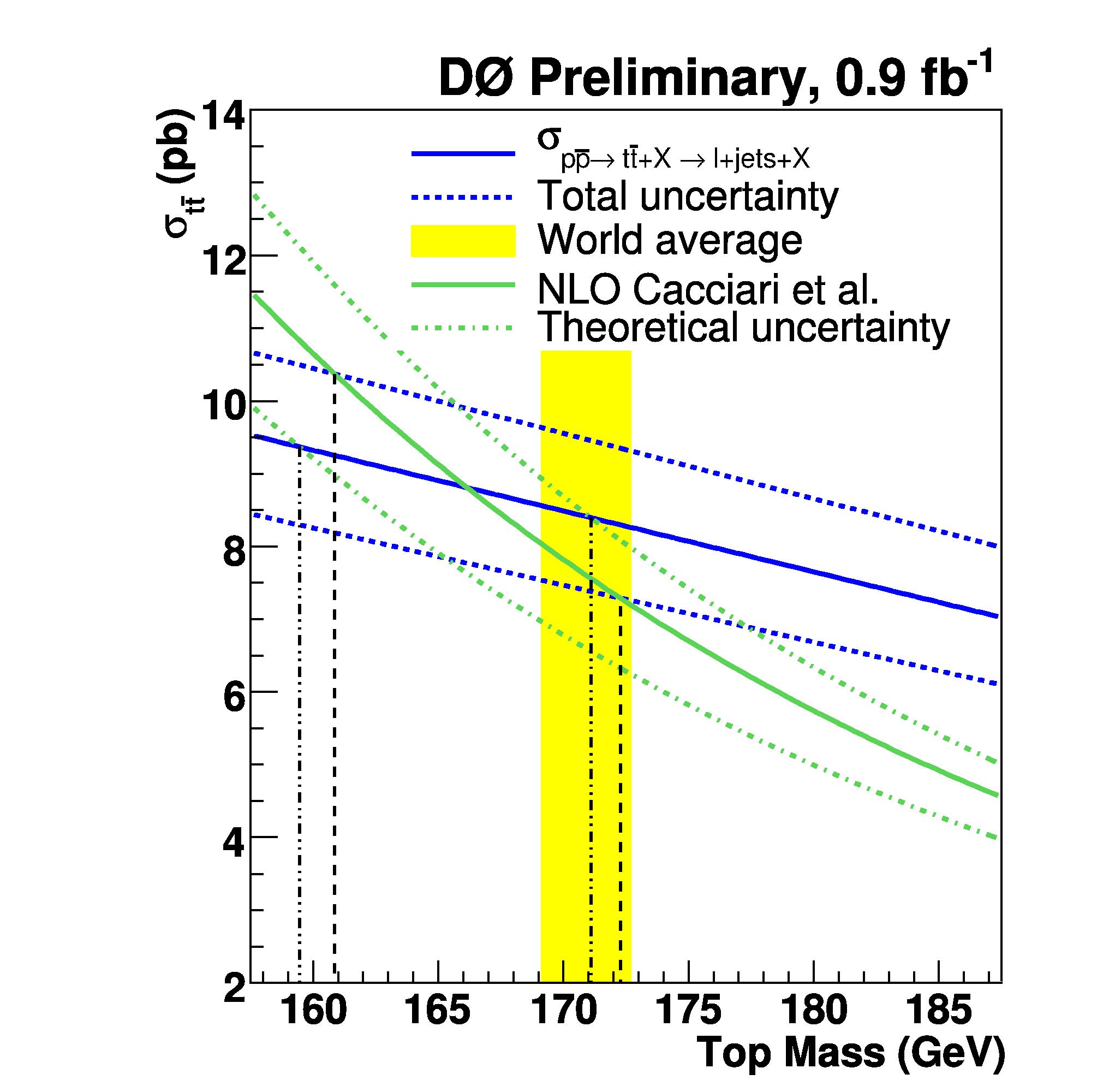}|\includegraphics[height=2in]{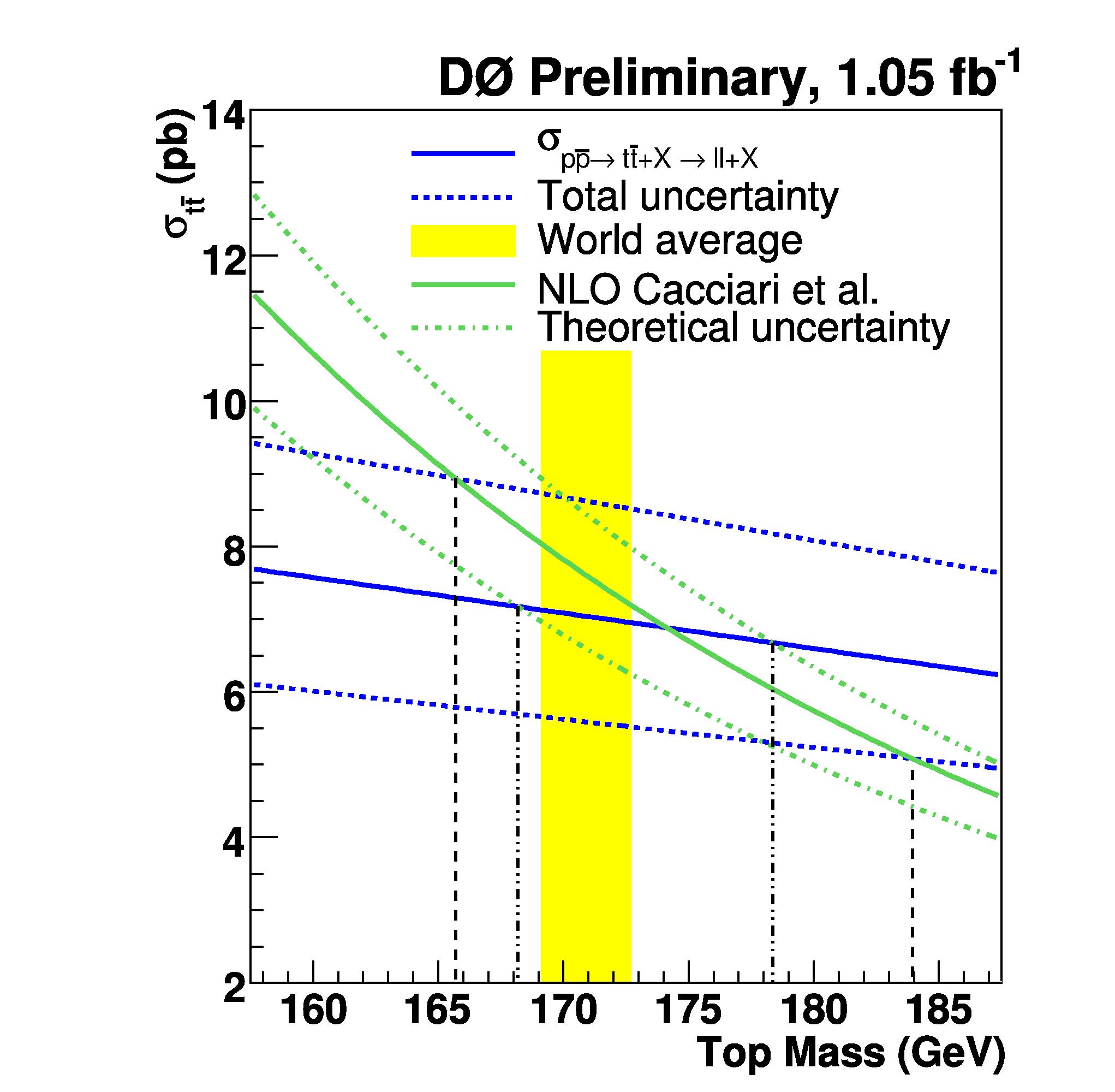}

\caption{Theoretical and experimental cross sections versus top mass for the lepton+jets and the dilepton channels.
\label{csmt}}
\end{figure}

This simple approach applied to the D0 measurements leads to the following values~\cite{deux} :
\begin{itemize}
\item for the lepton+jets channel : $M_{t}=166.1\ ^{+6.1}_{-5.3}$ (stat+sys) $ ^{+4.9}_{-6.7}$ (theory) GeV;

\item for the dilepton channel : $M_{t}=174.1\ ^{+9.8}_{-8.4}$ (stat+sys) $ ^{+4.2}_{-6.0}$ (theory) GeV.

\end{itemize}

Both values are in good agreement with the direct measurement.
\subsection{Second approach : use of probability rules}

Another approach is to understand the uncertainty bands of the figure 1 and to give some justified probability to all the points of these bands. So the different uncertainties on the cross section have to be determined and their shapes to be justified.

The quantity to evaluate is the probability density function $f(m_{t})$ for the top mass. As the top mass is extracted from cross section measurements, the uncertainties on the top mass are due to the uncertainties on the cross section. The proper knowledge and the combination of them allows to determine  $f(m_{t})$. The relation between the uncertainties on the cross section and the uncertainties on the top mass is given by :

\begin{equation}
 f(m_{t})=\int f(m_{t}|\sigma)f(\sigma)d\sigma
\end{equation}

\noindent where $f(\sigma)$ is the probability density function of having a given cross section, and it is taken to be flat as there is no preferred value for it.
Then the use of the Bayes theorem leads to :

\begin{equation}
 f(m_{t}|\sigma) =f(\sigma|m_{t})f_{0}(m_{t})
\end{equation}

\noindent where $f_{0}(m_{t})$ is the prior on $m_{t}$, taken to be flat. 

The quantity to evaluate is the probability density function for the cross section given the top mass, $f(\sigma|m_{t})$. At a given top mass, the different sources of uncertainties on the cross section are :
\begin{itemize}
\item experimental uncertainties~\footnote{The experimental cross section depends on the Parton Distribution Function such as the theoretical one. But to consider them independent, this contribution has been substract from the experimental uncertainties} : both systematic and statistic uncertainties lead to uncertainties in the expression of the measured cross section.
%\begin{equation}
 %  \sigma(p\bar{p}\rightarrow t\bar{t})=\frac{N_{observed}-N_{background}}{A_{tot}\int Ldt}
%\end{equation}
The shape of combined statistic and systematic uncertainty $ f_{exp}(\sigma|m_{t})$ is taken to be gaussian and both contributions are added quadratically.

\item theoretical uncertainties : the Parton Distribution Functions (PDF), the renormalization and factorization scale~\footnote{Both scales have been taken equal. For further studies, an independent variation could be considered.} uncertainties lead to an uncertainty band on the theoretical cross section given by the factorization theorem :

\begin{equation}
  \sigma_{tot}(p\bar{p}\rightarrow t\bar{t},S)=\sum_{i,j}\int dx_{i}dx_{j}f_{i,p}(x_{i},M)f_{j,\bar{p}}(x_{j},M)\hat{\sigma}_{i,j}(ij\rightarrow t\bar{t};\hat{s}=x_{i}x_{j}S,M)
\end{equation}
The shape of the PDFs uncertainty $f_{th,PDF}(\sigma|m_{t})$ is taken as gaussian. %with a width given by CTEQ :

%\begin{equation}
 %\Delta X=\frac{1}{2}\left( \sum_{i=1}^{N_{p}}\left[X(S_{i}^{+})-X(S_{i}^{-})\right]^{2}\right)^{1/2}
%\end{equation}

%where X is the observable (here the theoretical cross section), and $X(S_{i}^{\pm})$ are the predictions for X based on the PDF sets $S_{i}^{\pm}$ from the eigenvector basis. 
For the scale uncertainty $f_{th,\mu}(\sigma|m_{t})$, there is no well-defined shape : an interpretation of the uncertainty band on the theoretical cross section has to be assumed. A simple way is to take the uncertainty band at a $100\%$ confidence level (figure 2a), or $90\%$ confidence level and the remaining region~\footnote{For this region, a band twice larger as the former one has been defined and the parts that don't correspond to a $90\%$ confidence level are defined with a $10\%$ confidence level.} with a $10\%$ confidence level (figure 2b). A less hypothetical assumption is to use the theoretical relation between the cross section and the scales : this relation leads to prefer a higher value of the cross section due to a maximum at $\mu \sim m_{t}/2$~\cite{trois}. So the third possible shape (see figure 2c) is to put more weight on the high values of the cross section.
\end{itemize}

\begin{figure}[h!]
\centering
%\rule{5cm}{0.2mm}\hfill\rule{5cm}{0.2mm}
%\vskip 2.5cm
%\rule{5cm}{0.2mm}\hfill\rule{5cm}{0.2mm}

 \includegraphics[height=1.5in]{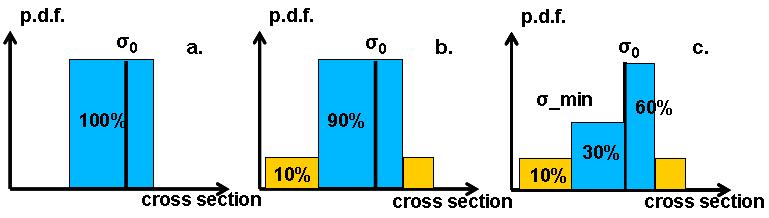}

\caption{Different shapes for the scale uncertainty : the probability density function (p.d.f.) illustrated as a function of the cross section.
\label{scale}}
\end{figure}

Finally, the probability density function of having a cross section at a given top mass is :
\begin{equation}
 f(m_{t}|\sigma)= f_{exp}(\sigma|m_{t})*(f_{th,PDF}(\sigma|m_{t})\otimes f_{th,\mu}(\sigma|m_{t}))
\end{equation}

\noindent where the experimental and theoretical uncertainties have been multiplied as they are considered independent and the theoretical uncertainties have been convoluted.

The method described above has been applied to the lepton+jets and dilepton channel measurements of the D0 collaboration. The central value for the top mass is not available yet but the uncertainty is about $\pm7$ GeV for the lepton+jets channel and $\pm10$ GeV for the dilepton channel, where the experimental uncertainty is bigger. Different choices for the theoretical uncertainty have limited impact on mass extraction (Variation of $\pm1$ GeV for the central values and the errors between the different choices).

\begin{figure}[h!]
\centering
%\rule{5cm}{0.2mm}\hfill\rule{5cm}{0.2mm}
%\vskip 2.5cm
%\rule{5cm}{0.2mm}\hfill\rule{5cm}{0.2mm}

 \includegraphics[height=2in]{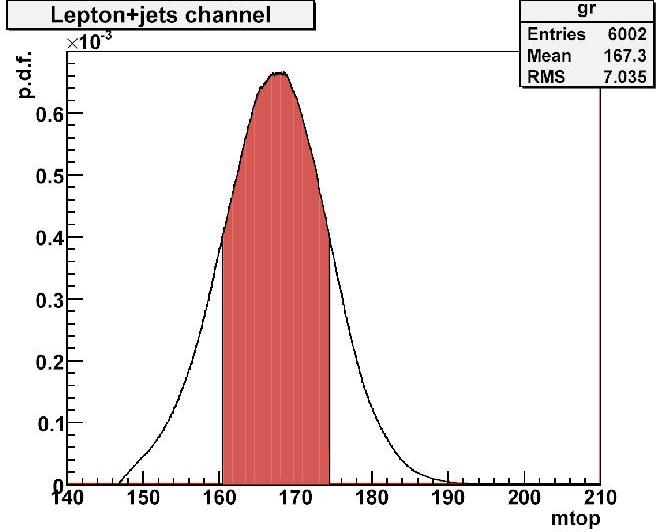}|\includegraphics[height=2in]{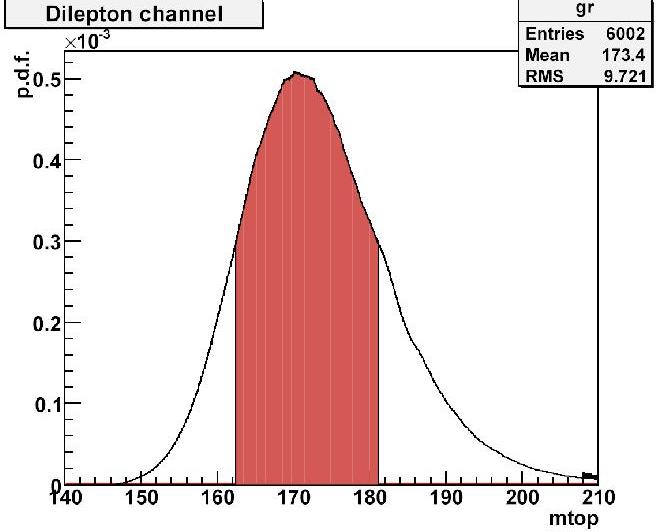}

\caption{The probability density function of the top mass for lepton+jets and the dilepton channels.
\label{result}}
\end{figure}

 \section{Summary}

 The extraction of the top mass from cross section measurements gives a value for the top mass in good agreement with the direct measurement but with a greater uncertainty. The uncertainties could be reduced by using measurements with a higher precision, new PDFs sets with smaller uncertainties or a more accurate higher order calculation. The extraction depends on the choice for the shape of the probability density function due to the scale uncertainties. The impact of choosing some other shapes than step functions will be investigated.

%\section*{Appendix}
 %We can insert an appendix here and place equations so that they are
%given numbers such as Eq.~\ref{eq:app}.
%\be
%x = y.
%\label{eq:app}
%\ee

\end{document}